\documentclass[aps,prd,twocolumn,letter]{revtex4}
\usepackage{ulem}
\usepackage{color}
\usepackage{graphicx}
\usepackage{epsfig}

\usepackage{bm}
\newcommand{\be}{\begin{equation}}
\newcommand{\ee}{\end{equation}}
\newcommand{\ba}{\begin{eqnarray}}
\newcommand{\ea}{\end{eqnarray}}

\begin{document}
\title{Toward a solution to the  $R_{AA}$ and $v_2$ puzzle for heavy quarks} 
\author{Santosh K. Das$^{a,b}$, Francesco Scardina$^{a,b}$, Salvatore Plumari$^{a,b}$, Vincenzo Greco$^{a,b}$}

\affiliation{$^a$ Department of Physics and Astronomy, University of Catania, 
Via S. Sofia 64, 1-95125 Catania, Italy}
\affiliation{$^b$ Laboratori Nazionali del Sud, INFN-LNS, Via S. Sofia 62, I-95123 Catania, Italy}

\date{}
\begin{abstract}
The heavy quarks constitutes a unique probe of the quark-gluon plasma properties.
Both at RHIC and LHC energies a puzzling relation between the nuclear modification 
factor $R_{AA}(p_T)$  and the elliptic flow $v_2(p_T)$ has been observed  which challenged
all the existing models, especially for D mesons.  We discuss how the temperature 
dependence of the heavy quark drag coefficient is responsible 
for a large part of such a puzzle. In particular, we have considered four different models 
to evaluate the temperature dependence of drag and diffusion coefficients 
propagating through a quark gluon plasma (QGP). All the 
four different models are set to reproduce the same $R_{AA}(p_T)$ observed in 
experiments at RHIC and LHC energy. We point out that for the same $R_{AA}(p_T)$ one can 
generate 2-3 times more $v_2$ depending on the temperature dependence of the heavy quark 
drag coefficient. A non-decreasing drag coefficient as $ T \rightarrow\ T_c \,$ is a major 
ingredient for a simultaneous description of $R_{AA}(p_T)$ and $v_2(p_T)$.

\vspace{2mm}

\end{abstract}
\maketitle

The ongoing nuclear collision programs at Relativistic Heavy Ion Collider (RHIC)
and Large Hadron Collider (LHC) energies have  created a medium that behaves 
like a nearly perfect fluid. The bulk properties of such a matter, called Quark Gluon Plasma (QGP),
 are governed by 
the light quarks and gluons ~\cite{Shuryak:2004cy,Science_Muller}. 
To characterize the QGP, penetrating and well calibrated probes are essential. In this context, 
the heavy quarks (HQs), mainly charm and bottom quarks, play a vital role since they do not constitute 
the bulk part of the matter owing to their larger mass compared to the temperature created in 
ultra-relativistic heavy-ion collisions (uRHIC's) \cite{hfr}. 

There are presently two main observables related to heavy quarks that have been measured at both
RHIC and LHC energies. The first one is the so-called nuclear suppression factor $R_{AA}$ that is the 
ratio between the $p_T$ spectra of heavy flavored hadrons (D and B) produced in nucleus + nucleus
collisions with respect to those produced in proton + proton collisions. More specifically at RHIC 
until recently has not been possible to measure directly D and B but only the leptons through 
their semileptonic decays. The other key observable is the elliptic flow $v_2=\langle cos(2\phi_p)\rangle$,
a measure of the anisotropy in the angular distribution that corresponds to the anisotropic 
emission of particles with respect to the azimuthal angle $\phi_p$. 
Despite their large mass, experimentally measured nuclear suppression factor $R_{AA}$ and 
elliptic flow $v_2$ of the heavy mesons are comparable to that of light hadrons~\cite{stare,phenixe,phenixelat,alice}.
This is in contrast to the expectations drawn initially from the perturbative interaction of HQs with 
the medium which predicted a $R_{AA} \approx 0.6$ for charm quarks, $R_{AA} \approx 0.8-0.9$ 
for bottom quarks in the central collisions \cite{Djordjevic:2005db,
Armesto:2005mz} at intermediate $p_T$. Also the  $v_2$ was predicted to be much smaller with respect to the
light hadron ones \cite{Armesto:2005mz}. 

Several theoretical efforts have been made 
in order to calculate the experimentally observed $R_{\mathrm AA}$ 
~\cite{stare,phenixe,phenixelat,alice} and $v_2$~\cite{phenixelat} for the non-photonic 
single electron spectra within the Fokker-Planck approach~\cite{DKS,moore,rappv2,rappprl,hiranov2,Gossiaux:2012ea,
alberico,bass,hees,bassnpa,he1,qun} and relativistic Boltzmann transport 
approach ~\cite{gossiauxv2,gre,Uphoff:2012gb,Younus:2013rja,Zhang:2005ni,Molnar:2006ci,Das:2013aga,our2,fs}.
Furthermore, also in a pQCD framework supplemented by the hard thermal loop
scheme several advances have been made to evaluate
realistic Debye mass and running coupling constants~\cite{gossiauxv2,alberico} 
and also three-body scattering effects~\cite{ko,Gossiaux:2012ea,Das,bass} have been implemented . 
It has been show in ~\cite{wicks} that the inclusion 
of both elastic and inelastic collisions  within a dynamical energy loss formalism 
reduces the gap between the theoretical and experimental results for 
$R_{AA}$ as $ p_T \geq 5-10\,$ GeV~\cite{Djo1,Djo2}. Several other improvements have been 
proposed  to advance the description of the data~\cite{adil,Djor,JA}.
Interaction from AdS/CFT~\cite{Gubser:2006qh} 
have also been implemented~\cite{hiranov2,horo,ali} to study the heavy flavor dynamics at RHIC and LHC.
Essentially all the models show some difficulties to describe simultaneously both 
$R_{AA}(p_T)$ and $v_2(p_T)$ and such a trait is not only present at RHIC energy but also 
in the results coming from collisions at LHC energy \cite{alice}.

In this letter we will address the impact of the temperature dependence of the interaction (drag coefficient) 
on  both $R_{AA}$ and $v_2$ relation. For this we are considering four different models having 
different T dependent drag coefficients. For the momentum evolution of the HQ, we are using 3+1 D 
Langevin dynamics. We notice that the several approaches and modelings of the HQ in-medium interaction 
differs significantly for the T dependence of the drag coefficient they entail. One can go from a 
$T^2$ dependence of the AdS/CFT approach to a drag coefficient that even increase with decrease T.
The aim of this letter is to show that, while generally a smaller $R_{AA}(p_T)$ corresponds 
to larger $v_2(p_T)$, the specific T dependence of the drag can strongly modify such an amount 
of $v_2(p_T)$, even if the models are tuned to reproduced the same $R_{AA}(p_T)$ observed experimentally. 
Our analysis shows that it is quite unlike that a drag with $T^2$ dependence can generate 
larger elliptic flow as the one observed experimentally at both RHIC and LHC.
Instead a nearly constant drag or an increasing one as $ T \rightarrow\ T_c \,$ strongly quenches the 
puzzling $R_{AA}(p_T)-v_2(p_T)$ relation.

The standard approach to HQ dynamics in the QGP is to follow their evolution by 
means of a Fokker-Plank equation solved stochastically by the Langevin equations.
The relativistic Langevin equations of motion for the evolution of the momentum 
and position of the heavy quarks can be written in the form
\begin{eqnarray}
 dx_i=\frac{p_i}{E×}dt, \nonumber \\
 dp_i=-\Gamma p_i dt+C_{ij}\rho_j\sqrt{dt}
 \label{lv1}
\end{eqnarray}
where $dx_i$ and $dp_i$ are the shift of the coordinate and momentum in each time step $dt$.
$\Gamma$ and $C_{ij}$ are the  drag force and the
covariance matrix in terms of independent Gaussian-normal 
distributed random variables $\rho$,$P(\rho)=(2\pi)^{-3/2}e^{-\rho^2/2}$, which obey 
the relations $<\rho_i \rho_j>=\delta_{ij}$ and $<\rho_i>=0$, respectively. 
The covariance matrix is related to the diffusion tensor, 
\begin{eqnarray}
C_{ij}=\sqrt{2B_0}P_{ij}^{\perp}+\sqrt{2B_1}P_{ij}^{\parallel}, 
\label{cmmm}
\end{eqnarray}
where $P_{ij}^{\perp}= \delta_{ij}-p_i p_j/p^2$ and $P_{ij}^{\parallel}=p_i p_j/p^2$ 
are the transverse and longitudinal projector operators respectively.
Under the assumption, $B_0=B_1=D$, 
Eq~(\ref{cmmm}) becomes $C_{ij}=\sqrt{2D(p)} \delta_{ij}$. Such an assumption strictly valid 
only for $p\rightarrow 0$, is usually employed at finite $p$ in application for heavy quark dynamics 
in the QGP \cite{moore,rappv2,bass,rappprl,Das,hees}.

We will discuss our results in terms of the drag coefficient $\Gamma$, but we remind that it is
related to the diffusion coefficient by the fluctuation-dissipation theorem that within a Langevin
approach reads $D=\Gamma E T$, for the case of the post-point Ito realization of the stochastic integral\cite{he-PRE}. In the post-point discretization the diffusion coefficients have to be used at the momentum argument p+dp, where dp is the increment from a pre-point Ito (Euler) time-step according to Eq.(\ref{lv1}). 

The solution of the stochastic Langevin equation needs a background medium describing the evolution of
the bulk QGP matter. To describe the
the expansion and cooling of the bulk and its elliptic flow $v_2(p_T)$ at both RHIC and LHC, we have employed a relativistic transport code with an initial condition given by a standard Glauber model and with an evolution at fixed $\eta/s=0.16$ (similarly to viscous hydro) 
see Ref.s \cite{Ruggieri:2013ova,Ruggieri:2013bda,Plumari:2011re,greco_cascade} for more details. 

Our objective is to demonstrate the effect of the temperature dependent interaction (drag coefficient)
on the $R_{AA}$ and $v_2$  obtained from different models. More specifically we investigate at fixed $R_{AA}$ how the $v_2$ is
 built up under various temperature dependence of the interaction.
For this purpose we consider four different modelings to calculate the 
drag and diffusion coefficients which are the key ingredients to solve the Langevin equation. 
Such models have to be considered merely as an expedient-device to generate different 
$T$ dependence of the $\Gamma(T)$ but the results and conclusions deduced will be 
much more general because they do not depend on the way the $\Gamma(T)$ has been obtained.
In this sense within a Fokker-Planck approach it is not relevant if the drag and 
diffusion coefficients has been evaluated considering only collisional or radiative loss.

\textit{Model-I (pQCD)} -
The elastic interaction of heavy quarks with the
light quarks, anti-quarks and gluons in the bulk has been considered within the
framework of pQCD to calculate the drag and diffusion coefficients. The scattering 
matrix ${\cal M}_{gHQ}$, ${\cal M}_{qHQ}$ and ${\cal M}_{\bar{q}HQ}$ 
are the well known Combridge matrix that includes $s,t,u$ channel and their interferences terms~\cite{comb}. 
The divergence associated with the $t$-channel diagrams due to massless intermediate
particle exchange has been shielded introducing the Debye screening mass $m_D=\sqrt{4\pi\alpha_s}\, T$. 
The temperature dependence of the coupling~\cite{zantow}:
\be
g^{-2}(T)=2\beta_0 ln\left(\frac{2\pi T}{a\,T_c ×}\right)+\frac{\beta_1}{\beta_0×}ln\left[ln\left(\frac{2\pi T}{a\,T_c ×}\right)\right]
\ee
where $\beta_0=(11-2N_f/3)/16\pi^2$, $\beta_1=(102-38N_f/3)/(16\pi^2)^2$ and $a=1.3$.
$N_f$ is the number of flavor and $T_C$ is the transition temperature.


\textit{Model-II (AdS/CFT)} -
We have also considered the drag force from the
gauge/string duality~\cite{Maldacena:1997re}, namely the conjectured equivalence between
conformal N=4 SYM gauge theory and gravitational theory in Anti de Sitter
space-time i.e. AdS/CFT. By matching the energy density of QCD and SYM, which leads 
to $T_{SYM}=T_{QCD}/{3^\frac{1}{4}}$, and the string prediction for quark-antiquark potential with lattice
gauge theory which gives $3.5<\lambda<8$ \cite{Gubser:2006qh}. 
One finds;
\be
\Gamma_{conf}= C \frac{T_{QCD}^2}{M_c} \\
\label{confdrag}
\ee
where $C={\pi\sqrt\lambda\over 2\sqrt3}=2.1\pm0.5$.
The corresponding diffusion constant D can be obtained from the
fluctuation-dissipation relation. Studies of heavy flavor momentum evolution 
within the Langevin dynamics using AdS/CFT can be found in Ref.~\cite{hiranov2,ali}.

\textit{Model-III (QPM)} -
The third model recently applied to estimate the heavy flavor transport coefficients 
is inspired by the quasi-particle model (QPM)~\cite{vc,elina,elina1}.
The QPM approach is a way to account for the non-perturbative dynamics
by T-dependent quasi-particle masses, $m_q=1/3g^2T^2$, $m_g=3/4g^2T^2$, plus 
a T-dependence background field known as bag constant.
Such an approach is able to successfully reproduce the thermodynamics of lQCD~\cite{salvo} 
by fitting the coupling $g(T)$.
To evaluate the drag and diffusion coefficients we have employed QPM tuned to the thermodynamics of 
the lattice QCD~\cite{lqcd_wb}. Such a fit lead to the following coupling~\cite{salvo}: 
\be
g^2(T)=\frac{48\pi^2}{[(11N_c-2N_f)ln[\lambda(\frac{T}{T_c}-\frac{T_s}{T_c})]^2}
\ee
where $\lambda$=2.6 and $T/T_s$=0.57.

\textit{Model-IV ($\alpha_{QPM}(T),m_q=m_g=0$)} - 
To have a different set of drag and diffusion coefficients we are considering a 
case where the light quarks and gluons are massless but the coupling is from 
the QPM which obtained from the fit to the lattice data. This case has to be 
mainly considered as an expedient to have a drag which decreasing with $T$ 
as obtained for example in the T-matrix approach~\cite{hfr,rappprl,he1}.

\begin{figure}[ht]
\begin{center} 
\includegraphics[width=17pc,clip=true]{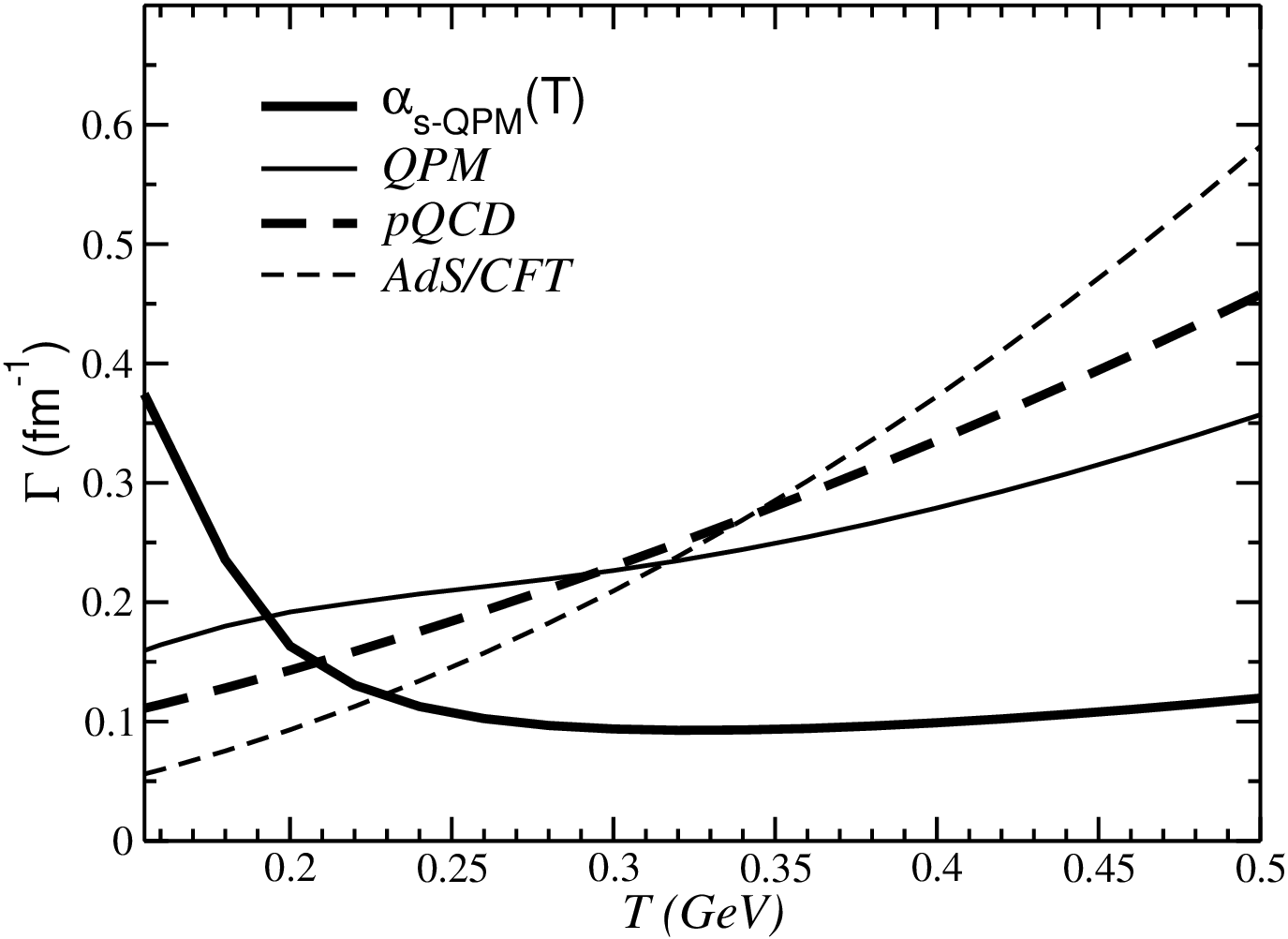}\hspace{2pc}
\caption{Variation of drag coefficient with respect to temperature.}
\label{fig1}
\end{center}
\end{figure}

In the following, except for the case of AdS/CFT,  we have calculated the drag coefficient
numerically from the scattering matrix of the model by means of the standard definition of drag 
\cite{BS}, see also Ref.\cite{elina1} for a recent detailed description of the calculation of the transport coefficients for heavy quarks.

The variation of the drag coefficient with respect to temperature at $p$=100 MeV obtained within the four 
different models discussed above has been shown in Fig.\ref{fig1}. The behaviors remain 
quite similar also at high momentum but with different magnitude. These rescaled drag coefficients 
can reproduce almost the same $R_{AA}$ at RHIC energy. In AdS/CFT case the drag coefficient is 
proportional to  $T^2$ whereas in $\alpha_{QPM}(T),m_q=m_g=0$ case the drag coefficient decrease with 
T due to the strong coupling at low temperature. It may be mentioned here that the drag coefficient 
obtained from the T-matrix~\cite{hfr,rappprl,riek} is almost constant or slightly decreasing with temperature. 

We mention  that the drag coefficient increases
with temperature when the system behaves like a gas. For a molecular liquid the drag
coefficient decreases with increasing temperature (except in a very few
cases) because a significant part of the thermal energy goes
into making the attraction between the interacting particles
weaker, allowing them to move more freely and hence reducing
the drag coefficient. The drag force of the partonic medium
with non-perturbative effects may decrease with increasing
temperature as shown in Ref.~\cite{hfr,rappprl,riek} because in this case the
medium interacts strongly more like a liquid. 

In order to study the impact of the temperature dependence of the drag coefficient
presented in the previous sections on the  experimental observables, we have 
calculated the nuclear suppression factor, $R_{AA}$,  using our initial 
charm and bottom quark distributions at initial time $t=\tau_i$ 
and final time $t=\tau_f$ at the freeze-out temperature  as $R_{AA}(p)=\frac{f(p,\tau_f)}{f(p,\tau_i)}$. 
Along with $R_{AA}$ we evaluate the anisotropic momentum distribution induced by 
the spatial anisotropy of the bulk medium and defined as

\be
 v_2=\left\langle  \frac{p_x^2 -p_y^2}{p_x^2+p_y^2}\right\rangle \ , \qquad \qquad
\ee
\label{eq5}

which measures the momentum space anisotropy.

\begin{figure}[ht]
\begin{center}
\includegraphics[width=17pc,clip=true]{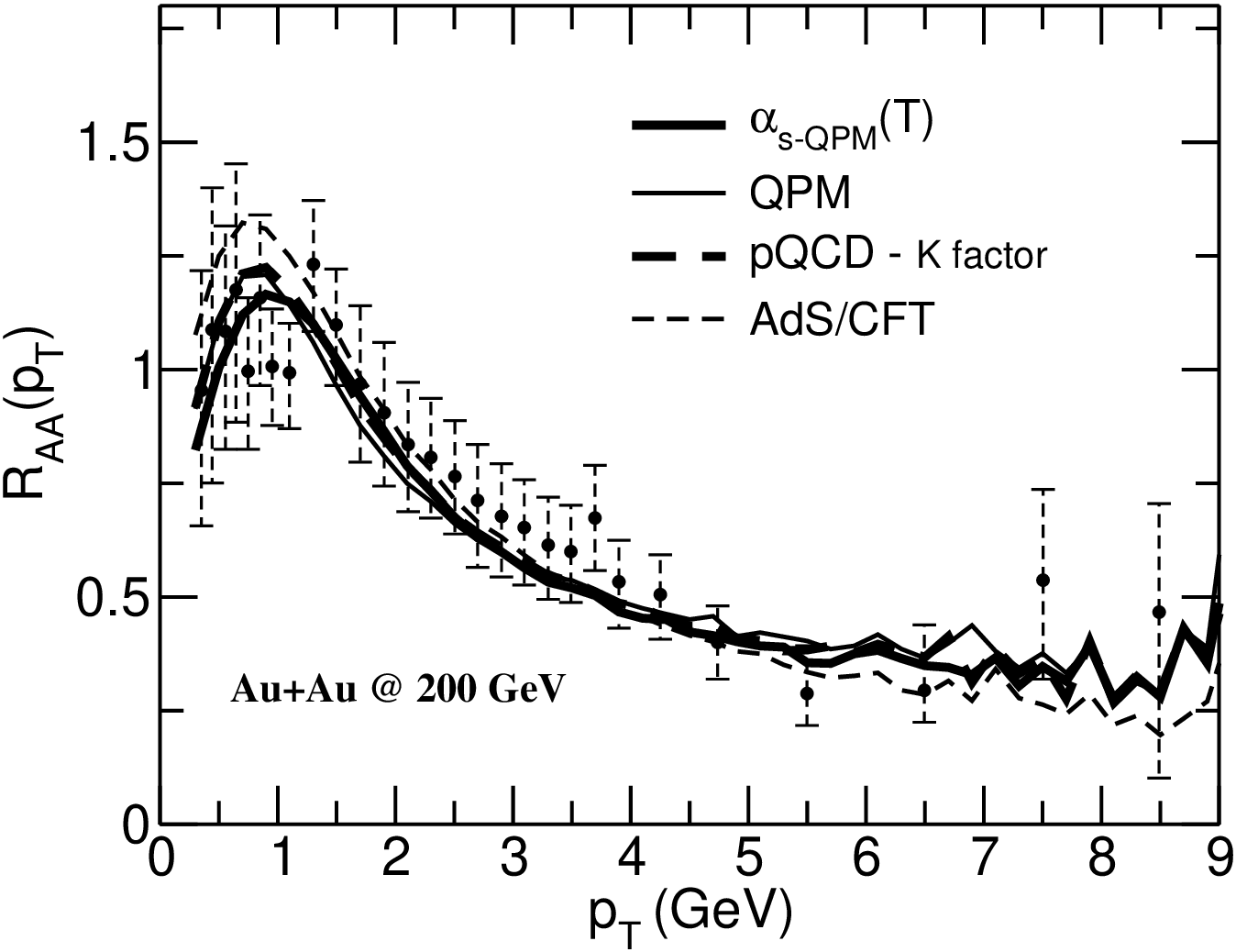}\hspace{2pc}
\caption{Comparison of the nuclear suppression factor, $R_{AA}$, as a function 
of $p_T$, obtained within the Langevin (LV) evolution for the four different cases, 
with the experimental data at RHIC energy.}
\label{fig2}
\end{center}
\end{figure}

\begin{figure}[ht]
\begin{center}
\includegraphics[width=17pc,clip=true]{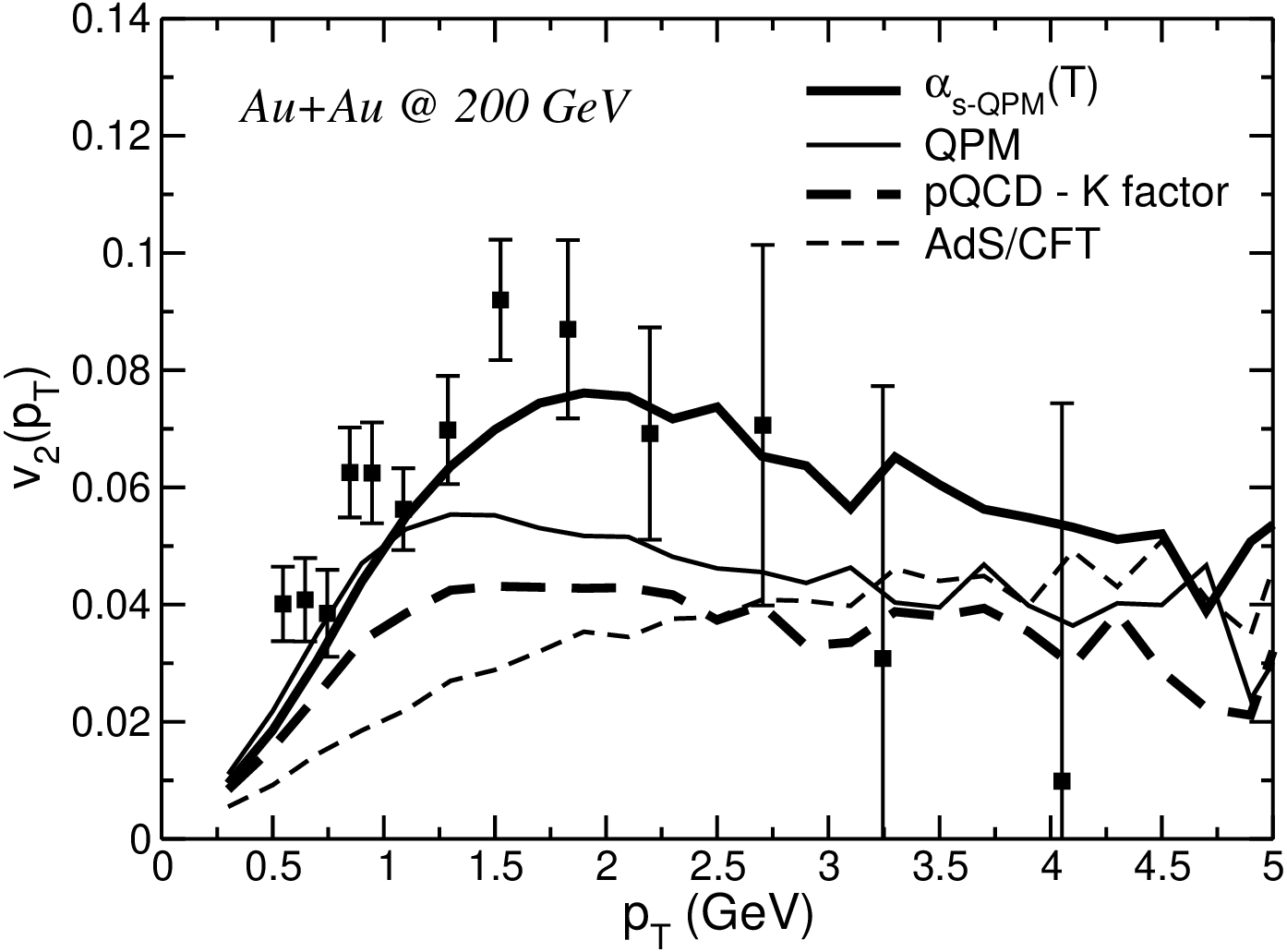}\hspace{2pc}
\caption{Comparison of the elliptic flow, $v_2$, as a function 
of $p_T$, obtained within the Langevin (LV) evolution for the four different cases, 
with the experimental data at RHIC energy.}
\label{fig3}
\end{center}
\end{figure}

\begin{figure}[ht]
\begin{center}
\includegraphics[width=17pc,clip=true]{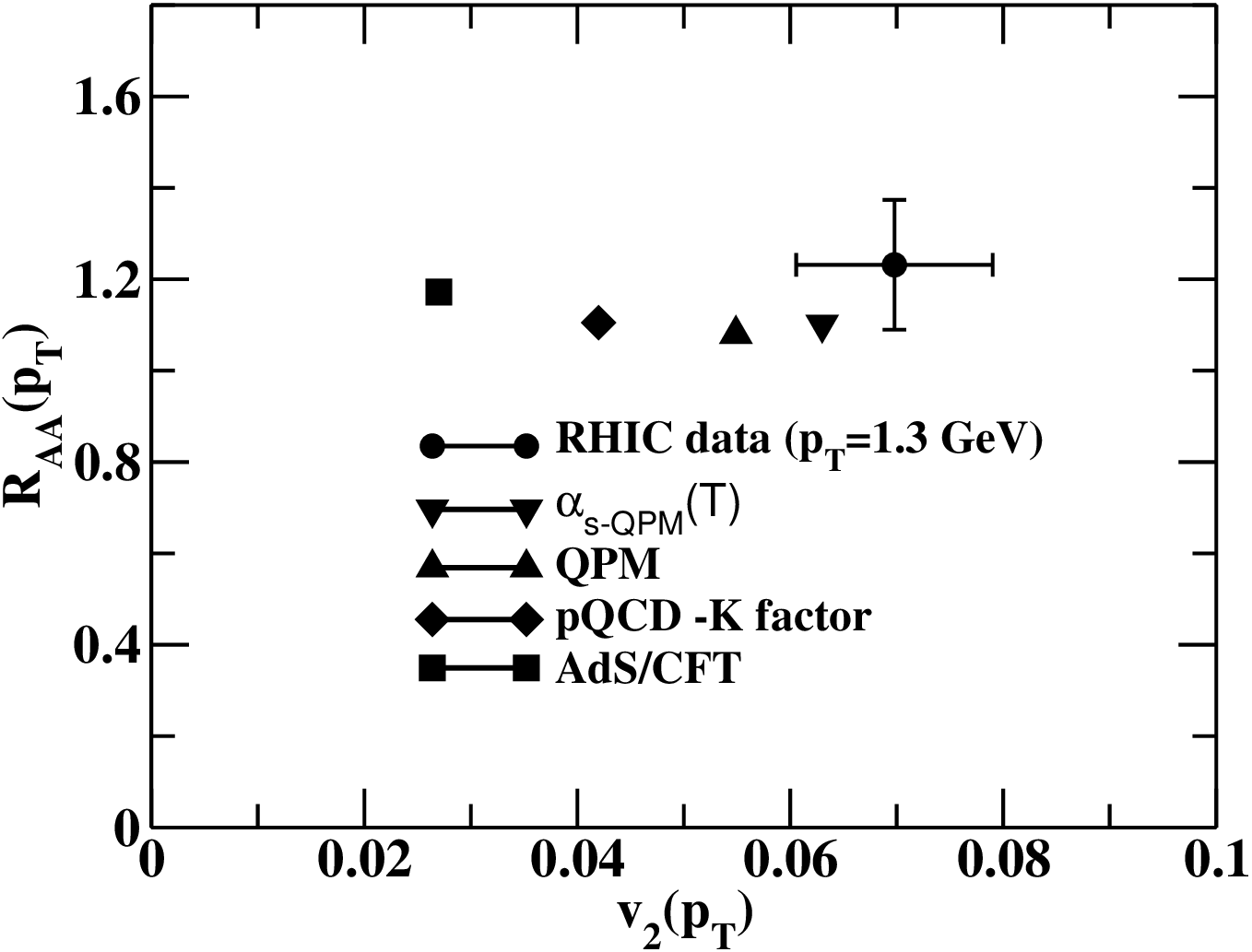}\hspace{2pc}
\caption{Comparison of the nuclear suppression factor $R_{AA}$  vs $v_2$, obtained 
within the Langevin (LV) evolution for the four different cases, 
with the experimental data at RHIC energy at $p_T=1.3$ GeV.}
\label{fig4}
\end{center}
\end{figure}

We have performed simulation of  $Au+Au$ collisions at $\sqrt{s}= 200$ AGeV
for the minimum bias using a 3+1D transport approach~\cite{greco_cascade,Ruggieri:2013bda,Scardina:2012hy}.
The initial conditions for the bulk evolution in the coordinate space are given by the  Glauber 
model condition, while in the momentum space we use a Boltzmann-Juttner distribution function 
up to a transverse momentum $p_T=2$ GeV and
at larger momenta mini-jet distributions as calculated within pQCD at NLO order \cite{Greco:2003xt}.
At RHIC energy, $Au+Au$ at $\sqrt{s}= 200$,
the maximum initial temperature of the fireball in the center is $T_i=340$ MeV and
the initial time for the fireball simulations is $\tau_i=0.6$ fm/c 
(according to the criteria $\tau_i \cdot T_i \sim 1$).  
The heavy quarks distribution in momentum space are distributed in accordance 
with the charm distribution in pp collisions that have been taken 
from ref~\cite{Cacciari:2005rk,Cacciari:2012ny} 
where in the coordinate space they are distributed according to $N_{coll}$.

The solution of the Langevin equation has been convoluted with the fragmentation functions 
of the heavy quarks at the quark-hadron transition temperature $T_c$ to 
obtain the momentum distribution of the D and B mesons.
For the fragmentation, we use the Peterson fragmentation function:
\be
f(z) \propto 
\frac{1}{\lbrack z \lbrack 1- \frac{1}{z}- \frac{\epsilon_c}{1-z} \rbrack^2 \rbrack}
\ee
where $\epsilon_c=0.04$ for charm quarks and  $\epsilon_c=0.005$ for bottom quark.

In Fig.~\ref{fig2} we have plotted $R_{AA}$ as a function of $p_T$ for the four 
different cases obtained within the Langevin dynamics at RHIC energy. As we mentioned, 
we try to reproduce the same $R_{AA}$ in all the cases by rescaling the drag and diffusion 
coefficients. 
We remind that RHIC data and calculations refer to the single electrons from the semileptonic
decay of D and B mesons.
The $v_2$ for the same $R_{AA}$ has been displayed in Fig~\ref{fig3} for all 
cases as a function of $p_T$. Our main striking point is that even if the $R_{AA}$ is very 
similar for all the four different cases, the $v_2$ built up is quite different depending on the 
temperature dependence of the drag coefficients (see Fig~\ref{fig1}). This is because the 
$R_{AA}$ is more sensitive to the early stage of the evolution whereas the $v_2$ 
is more sensitive to the later stage of the evolution (near $T_c$). Some  studies in this direction have been done also in the light flavor
sector as shown in Ref.~\cite{Liao:2008dk, Scardina:2010zz,liao1} and very recently related to
the presence of magnetic monopoles \cite{liao2}.
The larger drag coefficient is at low temperature the larger is the  $v_2$ even for the same $R_{AA}$. 
For example in the region of the peak for $v_2(p_T)$ we see a difference of about a factor 2.5 
going from a $T^2$ dependence, like AdS/CFT to a inverse T dependence as it can occur in a 
liquid. This last case or at least a nearly constant drag appears to be very much favored 
by the comparison with the data.

This study suggests that  the correct temperature dependence of drag coefficient has a
crucial role for a simultaneous reproduction of $R_{AA}$ and $v_2$. 
The reason for such a relation between the two observables is that a small $R_{AA}$
(strong suppression) can be generate very quickly at the beginning of the QGP lifetime,
i.e. at high T. However such a strong interaction will not be accompanied by a build-up
of $v_2$ because the bulk medium has not yet developed a sizeable part of its elliptic flow.
On the contrary to generate a large $v_2$ one needs that there is a strong interaction
with the medium at later stages of the QGP lifetime in order to match the build-up 
of both $R_{AA}$ and $v_2$. The experimental data seem to clearly suggest that the
drag of the medium cannot decrease with large power of T otherwise the interaction
will be relatively weak just when a strong interaction would make possible the
build of the anisotropy in momentum space.
It can be here mentioned 
that the drag coefficient is almost constant with respect to temperature in the T-matrix case~\cite{hfr,rappprl,he1,riek}. 
However also a QPM can be considered quite close to the data given that we have not included 
the coalescence mechanism that would shift the $v_2(p_T)$ in all the cases considered by about a 20-25$\%$ upward.
In Fig~\ref{fig4} we have introduced a new  plot $R_{AA}$ vs $v_2$ at a given momentum ($p_T$= 1.3 GeV) 
to promote the importance of simultaneous reproduction of $R_{AA}$ and $v_2$. Fig~\ref{fig4} highlights 
how the $v_2(p_T)$ built up can differ up to a factor of around 2.5 (in the region of peak), 
for the same $R_{AA}(p_T)$, depending on the temperature dependence of the drag coefficient.

\begin{figure}[ht]
\begin{center}
\includegraphics[width=17pc,clip=true]{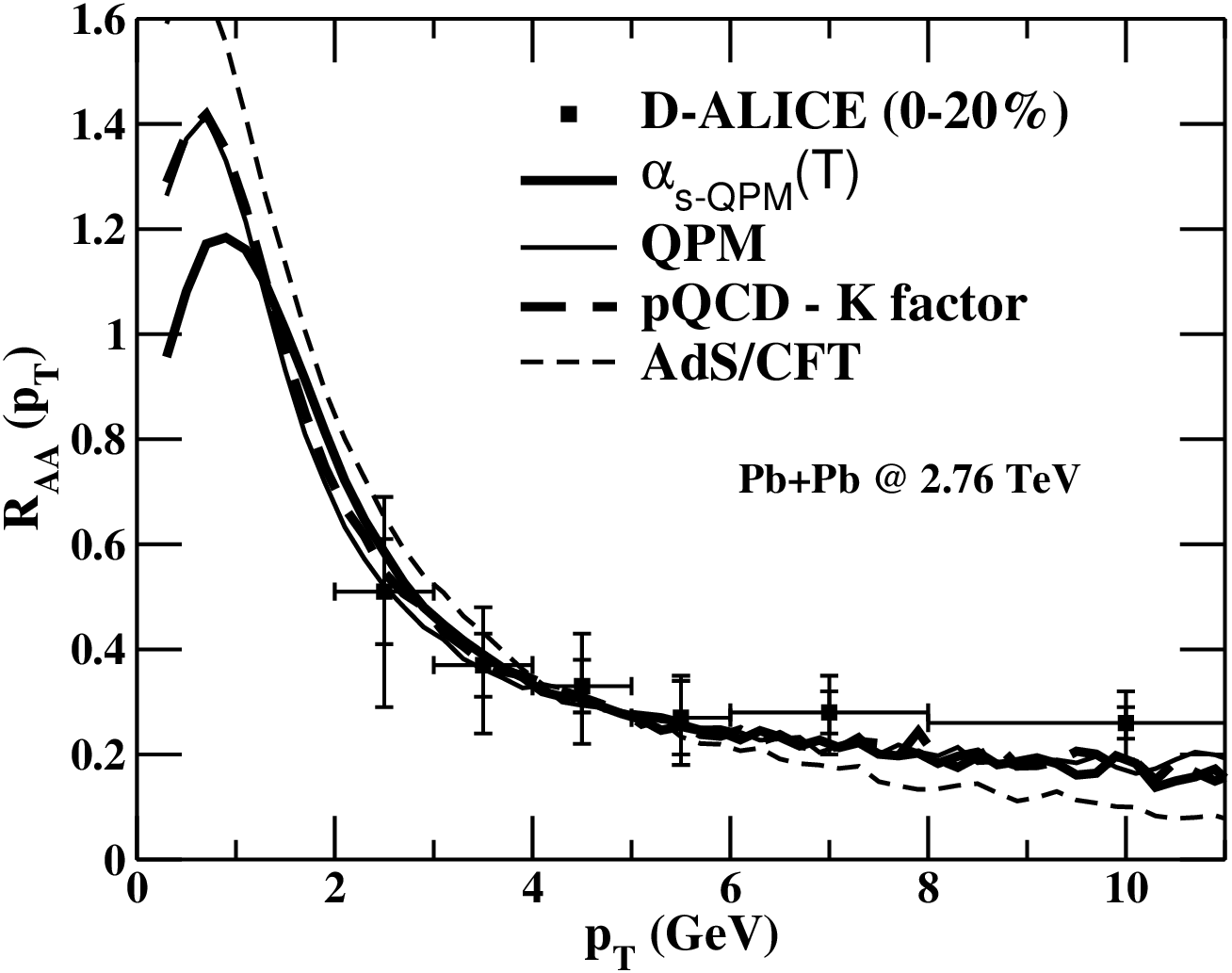}\hspace{2pc}
\caption{Comparison of the nuclear suppression factor, $R_{AA}$, as a function 
of $p_T$, obtained within the Langevin (LV) evolution for the four different cases, 
with the experimental data at LHC energy.}
\label{fig5}
\end{center}
\end{figure}

\begin{figure}[ht]
\begin{center}
\includegraphics[width=17pc,clip=true]{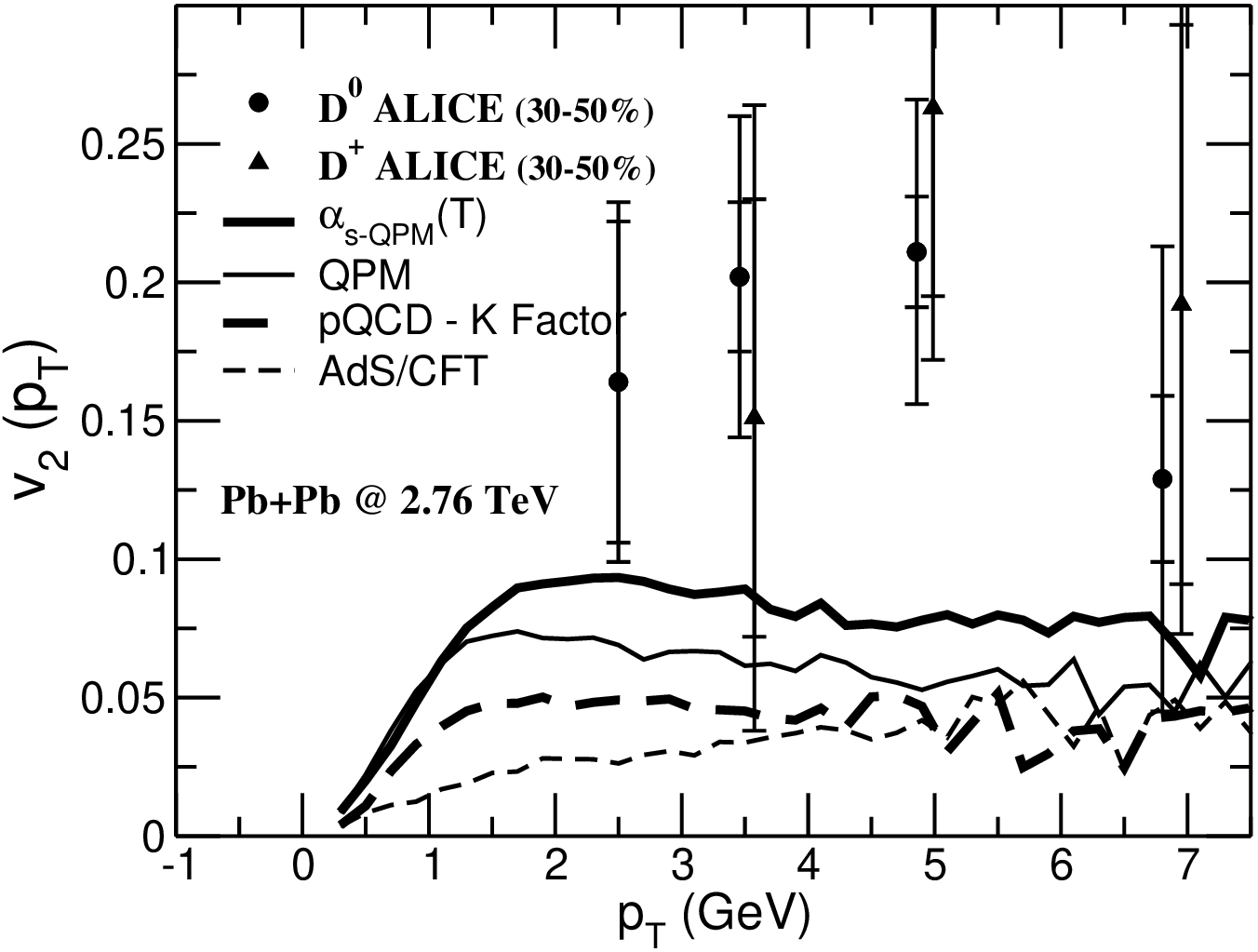}\hspace{2pc}
\caption{Comparison of the elliptic flow, $v_2$, as a function 
of $p_T$, obtained within the Langevin (LV) evolution for the four different cases, 
with the experimental data at LHC energy.}
\label{fig6}
\end{center}
\end{figure}

We have also extended our calculation to study $R_{AA}$ and $v_2$ at LHC performing 
simulations of  $Pb+Pb$  at $\sqrt{s}= 2.76$ ATeV energy. 
In this case the initial maximum temperature in the center of the fireball is $T_0=510$ MeV and
the initial time for the simulations is $\tau_0\sim 1/T_0 =0.3$ fm/c. 
In Fig.~\ref{fig5} we  show the $R_{AA}$ as a function of $p_T$ for the four 
different cases obtained within the Langevin dynamics at LHC energy. As we mentioned, 
we reproduce similar $R_{AA}$ in all the cases by rescaling the drag and diffusion 
coefficients. The elliptic flow $v_2$ for the same $R_{AA}$ has been plotted in 
Fig~\ref{fig6} for all cases as a function of $p_T$ . Similarly to the  RHIC case 
we get a similar trend for the $R_{AA}$ vs $v_2$ depending 
on the T dependence drag coefficients.

However, as pointed out in Ref~\cite{fs}, for charm quarks, which have  a moderate $M/T$ ratio, a 
significant deviation with respect to the Brownian Langevin dynamics can be expected. In this case
the full solution of the Boltzmann integral i.e. without the assumption of small collisional 
exchanged momenta, leads in general to a large $v_2(p_T)$. Such an effect depends on the anisotropy
of the microscopic scattering and can not be studied in term of only the drag coefficient. 
It is however an effect that in general can be expected to be of the order of about $20\%$ 
and does not modify the systematic studied here. A further effect that is involved in the study 
of HQ observable is related to the hadronization process. If the possibility of the 
coalescence process is included there is a further enhancement of the $v_2(p_T)$ 
of about a $20-25\%$ ~\cite{rappv2,rappprl,grecohq}. 
Also the hadronic rescattering may play a role in enhancing the $v_2(p_T)$ without 
modifying the $R_{AA}(p_T)$ \cite{He:2012xz}. 
This however would generate a similar shift for all the cases
discussed hence not affecting the discussed pattern entailed by $\Gamma(T)$.
The impact of Boltzmann dynamics 
and hadronization by coalescence are larger at LHC and can be lead to a better 
agreements with the data for the case $\alpha_{QPM}(T)$ and QPM but does not modify 
the impact of the T-dependence of the drag coefficient discussed in this letter.

The results shown have been obtained evaluating the drag $\Gamma$ from the respective models
and then the diffusion coefficient $D$ from FDT. Several other options are possible
like evaluating the diffusion from the scattering matrix and the drag from the FDT or employing
both drag and diffusion from the scattering matrix. We have seen that while these different
options may lead to some differences, once they are tuned to $R_{AA}$, the differences in the elliptic
flows stays within a $10\%$ and in particular our main result on the impact of the T dependence of the drag
is not affected by it.

In summary, we have evaluated the drag and diffusion coefficients of the heavy quarks 
within four different models.
With these transport coefficients and heavy quark initial distributions we have
solved the Langevin equation. The solution of Langevin equation has been
used to evaluate the  nuclear suppression factor, $R_{AA}$, and elliptic flow, $v_2$. 
The results have been compared with the experimental data both at RHIC and LHC energies.
Our primary intent is to highlight how the temperature dependence of the 
interaction (drag coefficient) provides  
an essential ingredient for the simultaneous 
reproduction of the  nuclear suppression factor, $R_{AA}$, and elliptic flow, $v_2$ 
which is a current challenge almost for all the existing model. 
Our work shows that the reproduction of the data on $R_{AA}(p_T)$ only cannot be 
used to determine the drag coefficient $\Gamma(T)$ of heavy quarks.
We find that the different T-dependences
of the drag coefficients in the literature can lead to differences in $v_2$ by 
2-3 times even if  the  $R_{AA}$ is very similar. Our study suggests the correct temperature 
dependence of the drag coefficient cannot be a large power of $T$, like $T^2$ as in pQCD or AdS/CFT.
We remind that $\Gamma(T)$ nearly constant or weakly decreasing with $T$ would 
be more typical of a liquid and not of a gas.


\vspace{2mm}
\section*{Acknowledgments}
We acknowledge the support by the ERC StG under the QGPDyn Grant n. 259684.

\end{document}